
\def\ung{{{\frak{g}}}}
\def\ungh{{{{\hat{\frak{g}}}}}}
\def\uqg{{{U_{q}(\ung)}}}
\def\uqgh{{{U_{q}(\ungh)}}}

\def\bp{{{\bold{P}}}}
\def\bq{{{\bold{Q}}}}
\def\calp{{{{\Cal{P}}}}}
\def\calq{{{{\Cal{Q}}}}}
\def\ot{{{\otimes}}}
\def\op{{{\oplus}}}

\def\unh{{{\frak{h}}}}

\input amstex
\documentstyle{amsppt}
\magnification 1200

\document
\centerline{\bf{{ Minimal Affinizations of Representations}}}
\vskip 12pt
\centerline{\bf of Quantum Groups:}
\vskip 12pt
\centerline{\bf the Rank 2 Case}
\vskip 36pt
\centerline{Vyjayanthi Chari{\footnote{Partially supported by the NSF,
DMS--9207701},}}
\vskip 12pt
\centerline{Department of Mathematics,}
\vskip 12pt
\centerline{University of California, Riverside, CA 92521.}
\vskip 36pt
\noindent{\bf Introduction}
\vskip12pt\noindent If $\ung$ is a finite--dimensional complex simple Lie
algebra, the associated \lq  untwisted\rq\  affine Lie algebra $\ungh$ is a
central extension, with one--dimensional centre, of the space of Laurent
polynomial maps $\Bbb C^{\times}\to\ung$ (on which a Lie bracket is defined
using pointwise operations). Since the cocycle of the extension vanishes on the
constant maps, we can regard $\ung$ as  a subalgebra of $\ungh$. If $V$ is any
representation of $\ung$, it is easy to extend the action of $\ung$ on $V$ to
an action of $\ungh$ on the same space. If $a\in\Bbb C^{\times}$, evaluation at
$a$ gives  a homomorphism $ev_a:\ungh\to\ung$ (under which the centre maps to
zero)
which is the identity on $\ung$, so pulling back $V$ by $ev_a$ gives the
desired extension. It follows from the results of [2] that, if $V$ is
finite--dimensional and irreducible, these are, up to isomorphism, the only
possible extensions.

Quantum deformations  $\uqg$ and $\uqgh$ of the universal enveloping algebras
of $\ung$ and $\ungh$ were introduced in 1985 by V. G. Drinfel'd and M. Jimbo.
These algebras depend on  a parameter $q\in\Bbb C^\times$; we  assume
throughout this paper that $q$ is transcendental. It is well--known (see [5] or
[9], for example) that, up to twisting by certain simple automorphisms, there
is a natural one-to-one correspondence between the finite--dimensional
representations of $\uqg$ and those of $\ung$. Corresponding representations
have the same character, and hence the same dimension. However, the structure
of the finite--dimensional representations of $\uqgh$ is not well--understood.
A parametrization of these representations in the spirit of Cartan's highest
weight classification of the finite-dimensional irreducible representations of
$\ung$ is proved in the case $\ung =sl_2$ in [3], and in [6] in general.

As in the classical situation, we may regard $\uqg$ as a subalgebra of $\uqgh$.
If $\ung$ is of type $sl_n$, the action of $\uqg$ on any representation $V$
extends to a representation of $\uqgh$. However, if $\ung$ is not of type
$sl_n$, it is not usually possible to extend the action of $\uqg$ on an
irreducible finite--dimensional representation $V$
to an action of $\uqgh$ on $V$.   Thus, it is natural to ask how $V$ can be \lq
enlarged\rq\ so as to obtain a representation of $\uqgh$. To make this question
precise, we define in this paper a natural partial ordering on the set of
isomorphism classes of representations of $\uqg$. By an {\it affinization} of a
finite-dimensional irreducible representation $V$  of $\uqg$, we mean an
irreducible representation $\hat V$ of $\uqgh$ which contains $V$ as a
$\uqg$-subrepresentation with multiplicity one, and such that all other
irreducible $\uqg$-subrepresentations of $\hat V$ are strictly smaller than
$V$. (There is a clear analogy with the classical Harish Chandra theory of
$(\ung,K)$-modules here.)

We prove that any given representation $V$ has only finitely many affinizations
(at least one) up to $\uqg$-isomorphism, and one may ask if any of them is \lq
canonical\rq. A reasonable interpretation of this question is to look for the
minimal affinization(s) of $V$, with respect to our partial order.
If $\ung=sl_n$, we show in [4] that every finite-dimensional irreducible
representation of $\uqg$ has, up to $\uqg$-isomorphism, a unique minimal
affinization. In this paper, we prove that, if $\ung$ is of type $C_2$ or
$G_2$, there is again a unique minimal affinization, and we describe it
precisely in terms of the highest weight classification of representations of
$\uqgh$ mentioned above. In contrast to the $sl_n$ case, the minimal
affinization in these cases is not, in general, irreducible as a representation
of $\uqg$. Subsequent papers will deal with the case when $\ung$ has rank
greater than 2.

The problem of constructing affinizations of representations of $\uqg$ is
important in several areas of mathematics and physics, as has been emphasized
by I. B. Frenkel and N. Yu. Reshetikhin, among others (see Remark 4.2 in [8]).
As one example, recall that, to any finite-dimensional irreducible
representation $V$ of $\uqg$ one can associate an R-matrix, i.e. an element
$R\in{\roman{End}}(V\ot V)$ which satisfies the \lq quantum Yang--Baxter
equation\rq\  (QYBE). There are many situations, however, in which it is
important to have a solution of the \lq QYBE with spectral parameters\rq\. This
is so, for example, in the theory of lattice models in statistical mechanics,
for only when the R-matrix constructed from the Boltzmann weights of the model
satisfies the QYBE with spectral parameters can one prove the existence of
commuting transfer matrices and deduce the integrability of the model. (See
[5], for example, for an introduction to these ideas.) Thus, it is natural to
ask when $R$ can be \lq embedded\rq\

in a parameter-dependent R-matrix $R(u)\in{\roman{End}}(V\ot V)$. A sufficient
condition for this is that the action of $\uqg$ on $V$ extends to an action of
$\uqgh$ on $V$, for then $V$ itself can be embedded in a 1-parameter family of
representations of $\uqgh$ by twisting with a certain 1-parameter family of
automorphisms of $\uqgh$ (which correspond, in the classical case, to \lq
rescaling\rq\ the $\Bbb C^\times$ parameter in $\hat\ung$).

A second example concerns the affine Toda field theory associated to
$\hat\ung$. This admits $U_q(\hat{\ung}^*)$ as a \lq quantum symmetry group\rq,
where $\hat{\ung}^*$ is the dual affine Lie algebra (whose generalized Cartan
matrix is the transpose of that of $\hat\ung$). It is well known that the
classical solitons of this theory correspond essentially to the
finite-dimensional irreducible representations of $\hat\ung$. The solitons (or
particle states) of the quantum theory should therefore correspond to the
finite-dimensional irreducible representations of $U_q(\hat{\ung}^*)$. Since
not all representations of $\uqg$ are affinizable on the same space, the
quantum solitons come in \lq multiplets\rq, and there are generally \lq
more\rq\ quantum solitons than classical ones.
\vskip 36pt

\noindent{\bf 1 Quantum affine algebras}
\vskip 12pt\noindent
Let $\ung$ be a finite--dimensional complex simple Lie algebra with Cartan
subalgebra $\unh$ and Cartan matrix $A= (a_{ij})_{i,j\in I}$. Fix coprime
positive integers  $(d_i)_{i\in I}$\/ such that $(d_ia_{ij})$\/ is symmetric.
Let $R$\/ be the set of roots and $R^+$\/ a set of positive roots. The roots
can be regarded as functions $I\to \Bbb Z$; in particular,
the simple roots $\alpha_i\in R^+$ are given by
$$\alpha_i(j) = a_{ji}, \ \ \ \ \ \  (i,j\in I).$$
Let $Q = \op_{i\in I}\Bbb Z.\alpha_i\subset\unh ^*$\/ be the root lattice, and
set $Q^+ =\sum_{i\in I}\Bbb N.\alpha_i$.

A weight is an arbitrary function $\lambda:I\to\Bbb Z$; denote the set of
weights by $P$, and let
$$P^+ =\{\lambda\in P: \lambda(i)\ge 0 \;\text{for all} \;i\in I\}$$
be the set of dominant weights. Define a partial order $\ge$\/ on $P$\/ by
$$\lambda\ge \mu \;\text{ iff}\; \lambda-\mu\in Q^+.$$
Let $\theta$\/ be the unique highest root with respect to $\ge$.

Define a non--degenerate symmetric bilinear form $(\ ,\ )$ on $\unh^*$ by
$$(\alpha_i,\alpha_j) = d_ia_{ij} ,$$
and denote by $(\ ,\ )$ also the induced form on $\unh$. Set
$d_0 =\frac12 (\theta,\theta)$, $a_{00} =2$, and, for all $i\in I$,
$$a_{0i} =-\frac{2(\theta,\alpha_i)}{(\theta, \theta)},\;\; a_{i0}
=-\frac{2(\theta, \alpha_i)}{(\alpha_i,\alpha_i)}.$$
Let $\hat{I} = I\amalg\{0\}$\/ and ${\hat A} =(a_{ij})_{i,j\in {\hat I}}$.
Then, ${\hat A}$\/ is  the generalized Cartan matrix of the untwisted affine
Lie algebra $\ungh$ associated to $\ung$.

Let $q\in \Bbb C^{\times}$ be transcendental, and, for $r,n\in\Bbb N$, $n\ge
r$, define
$$\align [n]_q & =\frac{q^n -q^{-n}}{q -q^{-1}},\\
[n]_q! &=[n]_q[n-1]_q\ldots [2]_q[1]_q,\\
\left[{n\atop r}\right]_q &= \frac{[n]_q!}{[r]_q![n-r]_q!}.\endalign$$
If $i\in\hat{I}$, let $q_i =q^{d_i}$.

\proclaim{Definition 1.1} With the above notation, $\uqgh$ is the unital
associative algebra over $\Bbb C$ with  generators $x_i^{{}\pm{}}$, $k_i^{{}\pm
1}$ ($i\in\hat I$), and the following defining relations:
$$\align
k_ik_i^{-1} = k_i^{-1}k_i &=1,\;\;  k_ik_j =k_jk_i,\\
k_ix_j^{{}\pm{}}k_i^{-1} &= q_i^{{}\pm a_{ij}}x_j^{{}\pm},\\
[x_i^+ , x_j^-] &= \delta_{ij}\frac{k_i - k_i^{-1}}{q_i -q_i^{-1}},\\
\sum_{r=0}^{1-a_{ij}}\left[{{1-a_{ij}}\atop r}\right]_{q_i}
(x_i^{{}\pm{}})^rx_j^{{}\pm{}}&(x_i^{{}\pm{}})^{1-a_{ij}-r} =0, \ \ \ \ i\ne
j.\tag 1\endalign$$

The algebra with generators $x_i^{{}\pm{}}$, $k_i^{{}\pm 1}$ ($i\in I$) and the
above defining relations (with the indices $i$, $j$\/ restricted to $I$) is
denoted by $\uqg$.
\endproclaim

Note that there is a canonical homomorphism of algebras $\uqg\to\uqgh$ which
takes $x_i^{{}\pm{}}\to x_i^{{}\pm{}}$, $k_i^{{}\pm 1}\to k_i^{{}\pm 1}$ for
all $i\in I$. The following result is well--known (see [5], for example).

\proclaim{Proposition 1.2} $\uqgh$ has the structure of a Hopf algebra, with
comultiplication $\Delta$, counit $\epsilon$, and antipode $S$, given by
$$\align\Delta(x_i^+)&= x_i^+\ot k_i +1\ot x_i^+,\\
\Delta(x_i^-)&= x_i^-\ot 1 +k_i^{-1}\ot x_i^-,\\
\Delta(k_i^{{}\pm 1}) &= k_i^{{}\pm 1}\ot k_i^{{}\pm 1},\\
\epsilon(x_i^{{}\pm{}}) =0,\;\ & \epsilon(k_i^{{}\pm 1}) =1,\\
S(x_i^+) = -x_i^+k_i^{-1},\; S(x_i^-) &=- k_ix_i^-, \; S(k_i^{{}\pm 1})
=k_i^{{}\mp 1},\endalign$$
for all $i\in\hat{I}$. Moreover, $\uqg$ is a Hopf algebra with structure maps
given by the same formulas, but with the index $i$ being restricted to the set
$I$. \qed\endproclaim

It is well--known that $\ungh$ may also be described as a central extension,
with one--dimensional centre, of the loop algebra of $\ung$, i.e. the space of
Laurent polynomial maps $\Bbb C^{\times}\to\ung$\/ under pointwise operations.
Drinfel'd [7] and Beck [1] give an analogous realization of $\uqgh$:

\proclaim{ Theorem 1.3}
Let $\Cal{A}_q$ be the unital associative algebra with generators
$x_{i,r}^{{}\pm{}}$ ($i\in I$, $r\in\Bbb Z$), $k_i^{{}\pm 1}$ ($i\in I$),
$h_{i,r}$ ($i\in I$, $r\in \Bbb Z\backslash\{0\}$) and $c^{{}\pm{1/2}}$, and
the following defining relations:
$$\align
c^{{}\pm{1/2}}\ &\text{are central,}\\
k_ik_i^{-1} = k_i^{-1}k_i =1,\;\; &c^{1/2}c^{-1/2} =c^{-1/2}c^{1/2} =1,\\
k_ik_j =k_jk_i,\;\; &k_ih_{j,r} =h_{j,r}k_i,\\
k_ix_{j,r}k_i^{-1} &= q_i^{{}\pm a_{ij}}x_{j,r}^{{}\pm{}},\\
[h_{i,r} , x_{j,s}^{{}\pm{}}] &= \pm\frac1r[ra_{ij}]_{q_i}c^{{}\mp
{|r|/2}}x_{j,r+s}^{{}\pm{}},\\
x_{i,r+1}^{{}\pm{}}x_{j,s}^{{}\pm{}} -q_i^{{}\pm
a_{ij}}x_{j,s}^{{}\pm{}}x_{i,r+1}^{{}\pm{}} &=q_i^{{}\pm
a_{ij}}x_{i,r}^{{}\pm{}}x_{j,s+1}^{{}\pm{}}
-x_{j,s+1}^{{}\pm{}}x_{i,r}^{{}\pm{}},\\
[x_{i,r}^+ , x_{j,s}^-]=\delta_{ij} & \frac{ c^{(r-s)/2}\phi_{i,r+s}^+ -
c^{-(r-s)/2} \phi_{i,r+s}^-}{q_i - q_i^{-1}},\\
\sum_{\pi\in\Sigma_m}\sum_{k=0}^m(-1)^k\left[{m\atop k}\right]_{q_i} x_{i,
r_{\pi(1)}}^{{}\pm{}}\ldots x_{i,r_{\pi(k)}}^{{}\pm{}} & x_{j,s}^{{}\pm{}}
 x_{i, r_{\pi(k+1)}}^{{}\pm{}}\ldots x_{i,r_{\pi(m)}}^{{}\pm{}} =0,\ \ i\ne j,
\endalign$$
for all sequences of integers $r_1,\ldots, r_m$, where $m =1-a_{ij}$,
$\Sigma_m$ is the symmetric group on $m$ letters, and the
$\phi_{i,r}^{{}\pm{}}$ are determined by equating powers of $u$ in the formal
power series
$$\sum_{r=0}^{\infty}\phi_{i,\pm r}^{{}\pm{}}u^{{}\pm r} = k_i^{{}\pm 1}
exp\left(\pm(q_i-q_i^{-1})\sum_{s=1}^{\infty}h_{i,\pm s} u^{{}\pm s}\right).$$

If $\theta =\sum_{i\in I}m_i\alpha_i$, set $k_{\theta} = \prod_{i\in
I}k_i^{m_i}$. Suppose that the root vector $\overline{x}_{\theta}^+$ of $\ung$
corresponding to $\theta$ is expressed in terms of the simple root vectors
$\overline{x}_i^+$ ($i\in I$) of $\ung$ as
$$\overline{x}_{\theta}^+ = \lambda[\overline{x}_{i_1}^+, [\overline
x_{i_2}^+,\cdots ,[\overline x_{i_k}^+, \overline x_j^+]\cdots ]]$$
for some $\lambda\in\Bbb C^{\times}$. Define maps $w_i^{{}\pm{}}:\uqgh\to\uqgh$
by
$$w_i^{{}\pm{}}(a) = x_{i,0}^{{}\pm{}}a - k_i^{{}\pm 1}ak_i^{{}\mp
1}x_{i,0}^{{}\pm{}}.$$
Then, there is an isomorphism of algebras $f:\uqgh\to\Cal A_q$ defined on
generators by
$$\align
f(k_0) = k_{\theta}^{-1}, \ f(k_i) &= k_i, \ f(x_i^{{}\pm{}}) =
x_{i,0}^{{}\pm{}},  \ \ \ \ (i\in I),\\
f(x_0^+) &=\mu w_{i_1}^-\cdots w_{i_k}^-(x_{j,1}^-)k_{\theta}^{-1},\\
f(x_0^-) &=\lambda k_{\theta} w_{i_1}^+\cdots w_{i_k}^+(x_{j,-1}^+),\endalign
$$
where $\mu\in\Bbb C^{\times}$ is determined by the condition
$$[x_0^+, x_0^-] =\frac{k_0-k_0^{-1}}{q_0-q_0^{-1}}. \qed$$
\endproclaim
Let $\hat U^{{}\pm{}}$ (resp. $\hat U^0$) be the subalgebra of  $\uqgh$
generated by the $x_{i,r}^{{}\pm{}}$ (resp. by the $\phi_{i,r}^{{}\pm{}}$) for
all $i\in I$, $r\in\Bbb Z$. Similarly, let $U^{{}\pm{}}$ (resp. $U^0$) be the
subalgebra of $\uqg$ generated by the $x_i^{{}\pm{}}$ (resp. by the $k_i^{{}\pm
1}$) for all $i\in I$. It is not difficult to prove
\proclaim{ Proposition 1.4} (a) $\uqg = U^-.U^0.U^+.$

(b) $\uqgh = \hat U^-.\hat U^0.\hat U^+.$ \qed
\endproclaim
It is clear that setting
$$deg(x_{i,r}^{{}\pm{}})=deg(h_{i,r})=r,\ \
deg(c^{{}\pm{1/2}})=deg(k_i^{{}\pm{1}})=0,\ \ \ (i\in I,\ r\in\Bbb Z)$$
gives $\uqgh$ the structure of a graded algebra. The following result is a more
precise formulation of this remark.
\proclaim{Proposition 1.5} For all $t\in\Bbb C^\times$, there exists a Hopf
algebra automorphism $\tau_t$ of $\uqgh$ such that
$$\align
\tau_t(x_{i,r}^{{}\pm{}})&=t^r(x_{i,r}^{{}\pm{}}),\ \
\tau_t(h_{i,r})=t^rh_{i,r},\\
\tau_t(k_i^{{}\pm{1}})&=k_i^{{}\pm{1}},\ \
\tau_t(c^{{}\pm{1/2}})=c^{{}\pm{1/2}}.
\endalign$$
\endproclaim
\demo{Proof} It is clear, as we have already said, that there is an algebra
automorphism $\tau_t$ given on generators by the above formulas. To see that
$\tau_t$ respects the coalgebra structure, note that, by the formula for the
isomorphism $f$ in 1.3,
$$\align
\tau_t(x_i^{{}\pm{}})&=x_i^{{}\pm{}},\ \ \tau_t(k_i^{{}\pm 1})=k_i^{{}\pm 1},\
\ (i\in I),\\
\tau_t(x_0^{{}\pm{}})&=t^{{}\mp{1}}x_0^{{}\pm{}},\ \ \tau_t(k_0^{{}\pm
1})=k_0^{{}\pm 1}.
\endalign$$
Using 1.2, it is easy to check that both sides of the equations
$$(\tau_t\ot\tau_t)\circ\Delta=\Delta\circ\tau_t,\ \ \tau_t\circ
S=S\circ\tau_t$$
agree on the generators in 1.1, and hence on the whole of $\uqgh$. \qed\enddemo
The $\tau_t$ are the quantum analogues of the \lq translation' automorphisms
which take a loop $\ell:\Bbb C^\times\to\ung$ to the loop $\ell_t$ given by
$\ell_t(u)=\ell(tu)$.

We shall also need to make use of the quantum analogue of the Cartan involution
of $\hat{\ung}$:
\proclaim{Proposition 1.6} There is a unique algebra involution $\hat\omega$ of
$\uqgh$ given on the generators of the presentation 1.3 by
$$\aligned
\hat\omega(x_{i,r}^{{}\pm{}})=-x_{i,-r}^{{}\mp{}},\ \
&\hat\omega(h_{i,r})=-h_{i,r},\\
\hat\omega(\phi_{i,r}^{{}\pm{}})=\phi_{i,-r}^{{}\mp{}},\ \
&\hat\omega(k_i^{{}\pm 1})=k_i^{{}\mp 1},\\
\hat\omega(c^{{}\pm{1/2}})&=c^{{}\mp{1/2}}.
\endaligned\tag2$$
Moreover, we have
$$\align
(\hat\omega\ot\hat\omega)\circ\Delta&=\Delta^{op}\circ\hat\omega,\tag3\\
\hat\omega^{-1}\circ S^{-1}\circ\hat\omega\circ S&=\kappa,\tag4\endalign$$
where $\Delta^{op}$ is the opposite comultiplication of $\uqgh$ and $\kappa$ is
the Hopf algebra automorphism of $\uqgh$ such that
$$\kappa(x_i^{{}\pm{}})=q_i^{{}\pm 2}x_i^{{}\pm{}},\ \ \kappa(k_i^{{}\pm
1})=k_i^{{}\pm 1}\ \ (i\in I).$$
\endproclaim
\demo{Proof} That the formulas in (2) do define an algebra involution of
$\uqgh$ is easily checked, using 1.3. To prove (3) and (4), we compute
$\hat\omega(x_0^{{}\pm{}})$. Note that, for any $a\in\uqgh$,
$$\align
\hat\omega(w_i^{{}\pm{}}(a))&=\hat\omega(x_{i,0}^{{}\pm{}}a-k_i^{{}\pm
1}ak_i^{{}\mp 1}x_{i,0}^{{}\pm{}})\\
&=-(x_{i,0}^{{}\mp{}}\hat\omega(a)-k_i^{{}\mp
1}\hat\omega(a)k_i^{{}\pm{}}x_{i,0}^{{}\mp{}})\\
&=-w_i^{{}\mp{}}(\hat\omega(a)).
\endalign$$
It follows from the formula for the isomorphism $f$ in 1.3 that
$$\align\hat\omega(x_0^+)&=-\mu(-1)^kw_{i_1}^+\ldots
w_{i_k}^+(x_{j,-1}^+)k_\theta\\
&=-\lambda^{-1}\mu(-1)^kq_0^2x_0^-,\endalign$$
and, because $\hat\omega$ is an involution,
$$\hat\omega(x_0^-)=-\lambda\mu^{-1}(-1)^kq_0^{-2}x_0^+.$$
Equations (3) and (4) are now easily checked on the generators in 1.1.
\qed\enddemo
It is clear that $\hat\omega$ is compatible, via the canonical map
$\uqg\to\uqgh$, with the Cartan involution $\omega$ of $\uqg$, given by
$$\omega(x_i^{{}\pm{}})=-x_i^{{}\mp{}},\ \ \omega(k_i^{{}\pm 1})=k_i^{{}\mp 1}\
\ \ (i\in I).$$

\vfill\eject

\vskip24pt\noindent{\bf 2 Finite--dimensional representations}
\vskip 12pt
\noindent Let $W$ be a representation of $\uqg$, i.e. a (left) $\uqg$--module.
One says that $\lambda\in P$\/ is  a weight of $W$ if the weight space
$$W_{\lambda} =\{w\in W\mid k_i.w = q_i^{\lambda(i)}w\}$$
is non--zero; the set of weights of $W$ is denoted by $P(W)$. We say that $W$
is of type 1 if
$$W =\bigoplus_{\lambda\in P(W)} W_{\lambda}.$$ The character of $W$ is the
function $ch_W: P\to\Bbb N$ given by $ch_W(\lambda) =dim (W_{\lambda})$.

If $W$ is a representation of $\uqg$, one says that $w\in W_{\lambda}$ is a
highest weight vector if $x_i^+.w =0$ for all $i\in I$. If $W =U_q(\ung).w$,
one says that $W$ is a highest weight representation with highest weight
$\lambda$. Lowest weight vectors and representations are defined similarly, by
replacing $x_i^+$ by $x_i^-$.

For a proof of the following proposition, see [5] or [9].
\proclaim{ Proposition 2.1} (a) Every finite--dimensional representation of
$\uqg$ is completely reducible.

(b) Every finite--dimensional irreducible representation of $\uqg$ can be
obtained from a type 1 representation by twisting with an automorphism of
$\uqg$.

(c) Every finite--dimensional irreducible representation of $\uqg$ of type 1 is
both highest and lowest weight. Assigning to such a representation its highest
weight defines a bijection between the set of isomorphism classes of
finite--dimensional irreducible representations of $\uqg$ and $P^+$.

(d) The finite--dimensional irreducible representation $V(\lambda)$ of $\uqg$
of highest weight $\lambda\in P^+$ has the same character as the irreducible
representation of $\ung$ of the same highest weight. \qed\endproclaim
By (a) and (c), if $W$ is any finite--dimensional representation of $\uqg$ of
type 1, we can write
$$W\cong\bigoplus_{\lambda\in P^+}V(\lambda)^{\op m_{\lambda}(W)}$$
for some uniquely determined multiplicities $m_{\lambda}(W)\in\Bbb N$. It will
be useful to define $m_{\lambda}(W) =0$ for $\lambda\in P\backslash  P^+$.
\proclaim{Proposition 2.1(continued)} (e) The multiplicities of the irreducible
components
in the tensor product $V(\lambda)\ot V(\mu)$, where $\lambda ,\mu\in P^+$, is
the same as in the tensor product of the irreducible representations of $\ung$
of the same highest weight.
\qed\endproclaim
 We now turn to representations of $\uqgh$. Note that, such a representation
$V$ may be regarded as a representation of $\uqg$ via the canonical
homomorphism $\uqg\to\uqgh$. We say that $V$ is of type 1 if $c^{1/2}$ acts as
the identity on $V$, and if $k_i$ acts semisimply on $V$ for all $i\in \hat I$.
Observe that $V$ is then of type 1 as a representation of $\uqg$; in
particular, the multiplicities $m_{\lambda}(V)$ ($\lambda\in P$) are
well--defined.

A vector $v\in V$ is a highest weight vector if
$$x_{i,r}^+.v=0,\ \ \phi_{i,r}^{{}\pm{}}.v=\Phi_{i,r}^{{}\pm{}}v,\ \ \ c^{1/2}.
v =v,$$
for some complex numbers $\Phi_{i,r}^{{}\pm{}}$. If, in addition, $V=\uqgh.v$,
then $V$ is called a highest weight representation, and the pair of
$(I\times\Bbb Z)$-tuples $(\Phi_{i,r}^{{}\pm{}})_{i\in I,r\in\Bbb Z}$ its
highest weight. Note that $\Phi_{i,r}^+=0$ (resp. $\Phi_{i,r}^-=0$) if $r<0$
(resp. if $r>0$), and that $\Phi_{i,0}^+\Phi_{i,0}^-=1$. (In [5], highest
weight representations of $\uqgh$ are called `pseudo-highest weight'.) Lowest
weight vectors and representations of $\uqgh$ are defined similarly.

The following result is proved in [5].
\proclaim{Proposition 2.2} (a) Every finite-dimensional irreducible
representation of $\uqgh$ can be obtained from a type 1 representation by
twisting with an automorphism of $\uqgh$.

(b) Every finite-dimensional irreducible representation of $\uqgh$ of type 1 is
both highest and lowest weight. \qed\endproclaim
Note, however, that in contrast to the case of $\uqg$, finite-dimensional
representations of $\uqgh$ are {\it not} completely reducible, in general.

The next result gives a parametrization of the finite-dimensional irreducible
representations of $\uqgh$ of type 1 analogous to that given for $\uqg$ by
2.1(c). If $\bp=(P_i)_{i\in I}$ is any $I$-tuple of polynomials $P_i\in\Bbb
C[u]$, its degree $deg(\bp)\in P^+$ is defined by
$$deg(\bp)(i)=deg(P_i).$$
Let $\Cal P$ be the set of $I$-tuples of polynomials with constant term 1, and,
for any $\lambda\in P^+$, let
$${\Cal P}^\lambda=\{\bp\in{\Cal P}\mid deg(\bp)=\lambda\}.$$
\proclaim{Theorem 2.3} Let $V$ be a finite-dimensional irreducible
representation of $\uqgh$ of type 1 and highest weight
$(\Phi_{i,r}^{{}\pm{}})_{i\in I,r\in\Bbb Z}$. Then, there exists
$\bp=(P_i)_{i\in I}\in\calp$ such that
$$\sum_{r=0}^\infty\Phi_{i,r}^+u^r=q_i^{deg(P_i)}\frac{P_i(q_i^{-2}u)}{P_i(u)}=\sum_{r=0}^\infty\Phi_{i,r}^-u^{-r},\tag5$$
in the sense that the left- and right-hand terms are the Laurent expansions of
the middle term about $0$ and $\infty$, respectively. Assigning to $V$ the
$I$-tuple $\bp$ defines a bijection between the set of isomorphism classes of
finite-dimensional irreducible representations of $\uqgh$ of type 1 and
$\calp$. \qed\endproclaim
This result is proved in [3] when $\ung=sl_2(\Bbb C)$, in [5] when
$\ung=sl_{n}(\Bbb C)$, and in [6] in the general case. We denote by $V(\bp)$
the finite-dimensional irreducible representation of $\uqgh$ associated to
$\bp\in\calp$. Abusing notation, we shall say that a representation $V$ as in
2.3 has highest weight $\bp$.

The next result describes the behaviour of the representations $V(\bp)$ under
tensor products. If $\bp=(P_i)_{i\in I}$, $\bq=(Q_i)_{i\in I}\in\calp$, let
$\bp\ot\bq\in\calp$ be the $I$-tuple $(P_iQ_i)_{i\in I}$. Obviously,
$deg(\bp\ot\bq)=deg(\bp)+deg(\bq)$.
\proclaim{Proposition 2.4} Let $\bp$, $\bq\in\calp$ be as above, and let
$v_{\bp}$ and $v_\bq$ be highest weight vectors of $V(\bp)$ and $V(\bq)$,
respectively. Then, in $V(\bp)\ot V(\bq)$,
$$x_{i,r}^+.(v_\bp\ot v_\bq)=0,\ \ \phi_{i,r}^{{}\pm{}}.(v_\bp\ot
v_{\bq})=\Psi_{i,r}^{{}\pm{}}(v_\bp\ot v_\bq),$$
where the complex numbers $\Psi_{i,r}^{{}\pm{}}$ are related to the polynomials
$P_iQ_i$ as the $\Phi_{i,r}^{{}\pm{}}$ are related to $P_i$ in (5).
\qed\endproclaim
The proof is essentially the same as that given in [3] when $\ung=sl_2(\Bbb
C)$.
\proclaim{Corollary 2.5} If $\bp$, $\bq\in\calp$, $V(\bp\ot\bq)$ is isomorphic
to a quotient of the subrepresentation of $V(\bp)\ot V(\bq)$ generated by the
tensor product of the highest weight vectors. \qed\endproclaim
Let $\lambda_i$ $(i\in I)$ be the fundamental weights of $\ung$:
$$\lambda_i(j)=\delta_{ij}\ \ \ \ \ \ (i,j\in I).$$
For any $a\in\Bbb C^\times$, let $V(\lambda_i,a)=V(\bp)$, where
$$P_j(u)=\left.\cases 1-a^{-1}u & \text{if $j=i$,}\\
1 & \text{if $j\ne i$.}\endcases\right .$$
The $V(\lambda_i,a)$ are called the fundamental representations of $\uqgh$.
This terminology is justified by the following consequence of 2.5:
\proclaim{Corollary 2.6} Every finite-dimensional irreducible representation of
$\uqgh$ of type 1 is isomorphic to a subquotient of a tensor product of
fundamental representations. \qed\endproclaim

\vskip36pt

\noindent{\bf 3 Minimal affinizations}
\vskip12pt\noindent
We propose the following definition.
\proclaim{Definition 3.1} If $\lambda\in P^+$, a finite-dimensional irreducible
representation $V$ of $\uqgh$ of type 1 is said to be an {\rm affinization} of
$\lambda$ if the highest weight $\bp$ of $V$ satisfies $deg(\bp)=\lambda$. Two
affinizations $V$ and $V'$ of $\lambda$ are said to be {\rm equivalent} if $V$
and $V'$ are isomorphic as representations of $\uqg$.\endproclaim
\vskip6pt\noindent{\it Remark 3.2.} It follows from 1.3 that, if $V$ is an
affinization of $\lambda$, then
$$V\cong V(\lambda)\oplus\bigoplus_{\{\mu\in
P^+\mid\mu<\lambda\}}V(\mu)^{m_\mu(V)}$$
as a representation of $\uqg$. Thus, $V$ gives a way of `extending' the action
of $\uqg$ on $V(\lambda)$ to an action of $\uqgh$, at the expense  of
`enlarging' $V(\lambda)$ by adding representations of $\uqg$ of smaller highest
weight.
\vskip12pt If $V$ is an affinization of $\lambda$, we denote its equivalence
class by $[V]$, and we write $\calq^\lambda$ for the set of equivalence classes
of affinizations of $\lambda$. Note that there is an obvious surjective map
$\calp^\lambda\to\calq^\lambda$, given by $\bp\mapsto[V(\bp)]$.

One can easily describe $\calq^\lambda$ in case $\lambda$ is fundamental:
\proclaim{Proposition 3.3} For any $i\in I$,
$$\calq^{\lambda_i}=\{[V(\lambda_i,1)]\}.$$\endproclaim
\demo{Proof} We need the following lemma, which will also be used elsewhere:
\proclaim{Lemma 3.4} Let $\rho:\uqgh\to End(V)$ be a finite-dimensional
irreducible representation of type 1 with highest weight $\bp=(P_i)_{i\in I}$.
For any $t\in\Bbb C^\times$, denote by $\tau_t^*(V)$ the representation
$\rho\circ\tau_t$. Then, $\tau_t^*(V)$ has highest weight $\bp^t=(P_i^t)_{i\in
I}$, where
$$P_i^t(u)=P_i(tu).$$
\endproclaim
\demo{Proof} This is immediate from (5), since
$$\tau_t(\phi_{i,r}^{{}\pm{}})=t^{{}\pm r}\phi_{i,r}^{{}\pm{}}\ \ \ \ \ (i\in
I,\ r\in\Bbb Z).\qed$$
\enddemo
If $V$ is an affinization of $\lambda_j$, with highest weight $\bp =
(P_i)_{i\in I}$, say, then $deg(P_i) = \lambda_j(i) = \delta_{ij}$, so $V\cong
V(\lambda_j, t)$\/ as representations of $\uqgh$, for some $t\in\Bbb
C^{\times}$. But then $\tau_t^*(V) \cong V(\lambda_j, 1)$. In particular,
$V\cong V(\lambda_j,1)$ as representations of $\uqg$. This proves 3.3.
\qed\enddemo

For arbitrary $\lambda\in P^+$, we have
\proclaim{Propostion 3.5} For any $\lambda\in P^+$, $\calq^{\lambda}$ is a
finite set.\endproclaim
\demo{Proof} Let $V$ be an affinization of $\lambda$, let
$\bp\in\Cal{P}^{\lambda}$ be the highest weight of $V$, and suppose that
$$P_i(u)=\prod_{r=1}^{\lambda(i)}(1 -a_{i,r}^{-1}u),$$
where $a_{i,r}\in\Bbb C^{\times}$. By 2.5, $V$ is isomorphic to a subquotient
of
$$\bigotimes_{i\in I}\left(\bigotimes_{r=1}^{\lambda(i)}V(\lambda_i,
a_{i,r})\right)$$
(the terms in the tensor products may be taken in any order). By 2.1(a) and
3.4, $V$ is isomorphic as a representation of $\uqg$ to a subrepresentation of
$$\bigotimes_{i\in I}V(\lambda_i,1)^{\ot\lambda(i)}.$$
Up to isomorphism, this representation obviously has only finitely many
subrepresentations, hence 3.5 is proved.\qed\enddemo
The rest of this section is devoted to the definition of a natural partial
order on $\Cal{Q}^{\lambda}$. It is convenient to first define a partial order
on a set which contains all of the $\Cal{Q}^{\lambda}$. Namely, if
$f:P^+\to\Bbb N$ is any function, let
$$supp(f)=\{\lambda\in P^+\mid f(\lambda) >0\}$$
and define
$$\Cal F =\{f\in\Bbb N^{P^+}\mid supp(f) \ \text{is finite}\}.$$
\proclaim{Definition 3.6} Let $f,g\in\Cal F$. We say that $f\preceq g$ iff, for
all $\mu\in P^+$, either

(i) $f(\mu)\le g(\mu)$, or

(ii) there exists $\nu >\mu$ with $f(\nu)< g(\nu)$.
\endproclaim
\proclaim{Proposition 3.7} $\preceq$ is a partial order on $\Cal F$.
\endproclaim
\demo{Proof} That $f\preceq f$, for all $f\in\Cal F$, is obvious. If $f\preceq
g$ and $g\preceq f$, suppose for a contradiction that there exists $\mu\in P^+$
with $f(\mu)\ne g(\mu)$. Note that,  since $supp(f)\cup supp(g)$ is finite,
there are at most finitely many such $\mu$, so we may assume that $\mu$ is
maximal (with respect to the partial order on $P^+$) among those weights for
which $f(\mu)\ne g(\mu)$. Without loss, assume that $f(\mu) < g(\mu)$. Since
$g\preceq f$, there exist $\nu >\mu$ with $f(\nu) >g(\nu)$; but this
contradicts the maximality of $\mu$.

Suppose finally that $f,g,h\in\Cal F$ are such that $f\preceq g$ and $g\preceq
h$, and assume for a contradiction that $f\preceq\!\!\!\!\!\!
\backslash \;h$. This means that there exists $\mu\in P^+$ such that
$$ f(\mu) >h(\mu) \;{\text{and, for all}} \ \nu >\mu,\ f(\nu)\ge h(\nu).\tag6$$
If 3.6(i) holds for $f$ and $g$, then by (6), $g(\mu) > h(\mu)$. Since
$g\preceq h$, there exists $\nu' >\mu$ with $g(\nu') <h(\nu')$. By (6),
$f(\nu')\ge h(\nu')$.
Thus,

(i)${}^\prime$ there exists $\nu'>\mu$ with $g(\nu') < h(\nu') \le f(\nu')$.

\noindent On the other hand, if (3.6)(ii) holds for $f$ and $g$, then, by (6),

(ii)${}^\prime$ there exists $\nu>\mu$ with $h(\nu)\le f(\nu) < g(\nu)$.

Note that any $\nu$ satisfying (ii)${}^\prime$ lies in $supp(g)$. Thus, if
there exists $\nu$ satisfying (ii)${}^\prime$, we may assume that $\nu$ is
maximal with this property. Since $g\preceq h$, there exists $\nu' >\nu$  with
$g(\nu') < h(\nu')$. Since $\nu'>\mu$, (6) implies that $\nu'$ satisfies
(i)${}^\prime$. But since $f\preceq g$, there exists $\nu'' >\nu'$ with
$f(\nu'') < g(\nu'')$. Then, (6) implies that $\nu''$ satisfies
(ii)${}^\prime$. Since $\nu'' >\nu$, this contradicts the maximality of $\nu$.

Similarly, assuming that (i)${}^\prime$ holds for some $\nu'$ also leads to a
contradiction.
\qed\enddemo
If $V$ is an affinization of $\lambda$, define $f_V\in\Cal F$ by
$$f_V(\mu) = m_{\mu}(V), \ \ \ \ \ \ (\mu\in P^+).$$
It is clear that $f_V$ depends only on the equivalence class of $V$, and that
the map $\Cal{Q}^{\lambda}\to\Cal F$ given by $[V]\to f_V$ is injective. Thus,
$\preceq$ induces a partial order on $\Cal Q^{\lambda}$, which we also denote
by $\preceq$.
\proclaim{Defintion 3.8} If $\lambda\in P^+$ and $[V]$,
$[W]\in\Cal{Q}^{\lambda}$, we write $[V]\preceq [W]$ iff, for all $\mu\in P^+$,
either,

(i) $m_{\mu}(V)\le m_{\mu}(W)$, or

(ii) there exists $\nu>\mu$ with $m_{\nu}(V) < m_{\nu}(W)$.

An affinization $V$ of $\lambda$ is {\rm minimal} if $[V]$ is a minimal element
of $\Cal{Q}^{\lambda}$ for the partial order $\preceq$, i.e. if
$[W]\in\Cal{Q}^{\lambda}$ and $[W]\preceq [V]$ implies that $[V]=[W]$.
\endproclaim
It follows immediately from 3.5 that
\proclaim{Corollary 3.9} For any $\lambda\in P^+$, minimal affinizations of
$\lambda$ exist. \qed
\endproclaim

\vskip24pt\noindent{\bf 4 The rank 1 case}
\vskip12pt\noindent
In this section, $\ung = sl_2(\Bbb C)$ and $I =\{1\}$.
\proclaim{ Defintion 4.1} Let $r\in\Bbb N$, $a\in\Bbb C^{\times}$. The $q$-{\rm
segment} $S_{r,a}$ of length $r$ and centre $a$ is the sequence of non--zero
complex numbers $aq^{-r+1}, aq^{-r+3},\cdots , aq^{r-1}$. Two $q$--segments
$S_1$ and $S_2$, of lengths $r_1$ and $r_2$, are  said to be in {\rm special
position} if $S_1\cup S_2$ is, when suitably ordered, a $q$-segment of length
strictly greater than $max\{r_1,r_2\}$; otherwise, $S_1$ and $S_2$ are said to
be in {\rm general position}.
\endproclaim
We recall the main result of [3]:
\proclaim{Theorem 4.2} Let $r\in\Bbb N$, $a\in\Bbb C^{\times}$ and write
$$P_{r,a}(u) = \prod_{k=1}^r(1-a^{-1}q^{r-2k+1}u),$$
so that the roots of $P_{r,a}$ are the elements of $S_{r,a}$. Then:

(a) $V(P_{r,a})$ is irreducible as  a representation of $U_q(sl_2)$, and has
dimension $r+1$;

(b) a tensor product

$$ V(P_{r_1, a_1})\ot V(P_{r_2,a_2})\ot\cdots \ot V(P_{r_m,a_m}), \tag7$$
where $r_1,r_2,\ldots , r_m\in\Bbb N$, $a_1,a_2,\ldots a_m\in\Bbb C^{\times}$,
is irreducible as a representation of $U_q(\hat{sl}_2)$ iff each pair of
strings $S_{r_k,a_k}$, $S_{r_l,a_l}$, for $1\le k < l\le m$, is in general
position. Moreover, two irreducible tensor products of the form (7) are
isomorphic as representations of $U_q(\hat{sl}_2)$ iff one is obtained from the
other  by permuting the factors in the tensor product;

(c) every finite--dimensional irreducible representation of $U_q(\hat{sl}_2)$
of type 1 is isomorphic to a tensor product of the form (7).\qed\endproclaim
\proclaim{Corollary 4.3} For any $r\in\Bbb N$, $\Cal{Q}^{r\lambda_1}$ has a
unique minimal element. This element is represented by $V(P)$, where $P$ is any
polynomial of degree $r$ whose roots form a $q$--segment.
If $[W]\in\Cal{Q}^{r{\lambda_1}}$ is not minimal, then $m_{(r-2)\lambda_1}\!(W)
> 0$.\endproclaim
\demo{Proof} The first part is immediate from 4.2 and 3.4. If
$[W]\in\Cal{Q}^{r\lambda_1}$ is not minimal, then by 4.2(c),
$$W\cong V(P_{r_1,a_1})\ot\cdots\ot V(P_{r_m,a_m})$$ where $r_1+\cdots +r_m =r$
and $m >1$. By 2.1(e) and the well--known Clebsch--Gordan decomposition for
representations of $sl_2(\Bbb C)$, the second part of 4.3 follows.\qed\enddemo

We record the following result here, as it will be needed later. It is an
immediate consequence of Propsition 4.9 in [3].
\proclaim{Proposition 4.4} Let $r,s\in\Bbb N$, $a,b\in\Bbb C^{\times}$, and let
$v,w$ be $U_q(\hat{sl}_2)$ highest weight vectors in $V(P_{r,a})$,
$V(P_{s,b})$, respectively. Then, $W = U_q(\hat{sl}_2).(v\ot w)$ is a proper
$U_q(\hat{sl}_2)$--subrepresentation of $V = V(P_{r,a})\ot V(P_{s,b})$ iff
$b/a =q^{r+s-2p+2}$ for some $0 <p \le min\{r,s\}$. In that case, $W$ and $V/W$
are irreducible as representations of $U_q(\hat{sl}_2)$, and as representations
of $U_q(sl_2)$,
$$\align W&\cong V((r+s)\lambda_1)\op V((r+s-2)\lambda_1)\op \cdots \op
V((r+s-2p+2)\lambda_1),\\
V/W&\cong V((r+s-2p)\lambda_1)\op V((r+s-2p-2)\lambda_1)\op\cdots\op
V(|r-s|\lambda_1).\qed\endalign$$\endproclaim

\vskip24pt\noindent{\bf 5 The rank 2 case}
\vskip12pt\noindent In this section, $\ung$ is of rank 2 and $I=\{1,2\}$.
\proclaim{Theorem 5.1} Let $\lambda=r_1\lambda_1+r_2\lambda_2\in P^+$. Then,
$\calq^\lambda$ has a unique minimal element. This element is represented by
$V(\bp)$, where $\bp\in\calp^\lambda$, iff the following two conditions are
satisfied:

(a) for each $i=1$, $2$, either $P_i=1$ or the roots of $P_i$ form a
$q_i$-segment of length $r_i$ and centre $a_i$ (say);

(b) if $P_1\ne 1$ and $P_2\ne 1$, then
$$\frac{a_1}{a_2}=q^{d_1r_1+d_2r_2+2d_2-1}\ \ \text{or}\ \
q^{-(d_1r_1+d_2r_2+2d_1-1)}.$$
\endproclaim
The proof of 5.1 will occupy the remainder of this section. If $\ung$ is of
type $A_2$, it is proved in [4]. From now on, we assume that $\ung$ is of type
$C_2$ or $G_2$.
\vskip 8pt
\noindent{\it Proof of 5.1(a).} To prove that 5.1(a) is necessary, we need the
following two lemmas. To state the first lemma, we note  that, for each $i=1$,
2, there is an algebra homomorphism $U_{q_i}(\hat{sl}_2)\to\uqgh$ such that
$x_{1,r}^{{}\pm{}}\mapsto x_{i,r}^{{}\pm{}}$, $k_1\mapsto k_i$, $h_{1,r}\mapsto
h_{i,r}$ (this is clear from 1.3). Let $\hat{U}_i$ be the image of this map.
\proclaim{Lemma 5.2} Let $\bp\in\calp$ and let $v_\bp$ be a highest weight
vector of $V(\bp)$. Then, for each $i=1$, $2$, $\hat{U}_i.v_\bp$ is an
irreducible representation of $U_{q_i}(\hat{sl}_2)$.\endproclaim
\proclaim{Lemma 5.3} Let $\lambda\in P^+$, $\bp\in\calp^\lambda$.

(a) If $P_i\ne 1$ and the roots of $P_i$ do not form a $q_i$-segment, then
$m_{\lambda-\alpha_i}(V(\bp))>0$.

(b) If $P_i=1$, or if $P_i\ne 1$ and the roots of $P_i$ form a $q_i$-segment,
then, for all $r>0$, $m_{\lambda-r\alpha_i}(V(\bp))=0$.\endproclaim

Assuming these lemmas, suppose that $V(\bp)$ is minimal but that, for some
$i\in I$, $P_i\ne 1$ and the roots of $P_i$ do not form a $q_i$-segment. Let
$\bq=(Q_j)_{j\in I}\in\calp^\lambda$ be such that, for $j=1,2$,  if $Q_j \ne 1$
the roots of $Q_j$ form a $q_j$-segment. We claim that $[V(\bq)]\prec[V(\bp)]$.

Let $\mu\in P^+$ be such that $m_\mu(V(\bq))>0$, $\mu\ne\lambda$. By 5.3(b),
$\mu=\lambda-s_1\alpha_1-s_2\alpha_2$, where $s_1$, $s_2>0$. Hence,
$\mu<\lambda-\alpha_i$, and, by 5.3(a),
$$m_{\lambda-\alpha_i}\!(V(\bp))>0,\ \ m_{\lambda-\alpha_i}\!(V(\bq))=0,$$
so $\mu$ satisfies 3.8(ii). This proves our claim, and hence also that 5.1(a)
is necessary.
\vskip8pt
\noindent\demo{ Proof of 5.2} Suppose that $\hat{U}_i.v_{\bp}$ is reducible.
Then, by 2.2(b), there exists $v\in\hat{U}_i.v_{\bp}$, not a multiple of
$v_{\bp}$, such that $v$ is annihilated by $x_{i,r}^+$ for all $r\in\Bbb Z$ and
is an eigenvector of $k_i$.
It is easy to see from the relations in 1.3 that the set of such vectors $v$ is
preserved by the action of the $\phi_{j,s}^{{}\pm{}}$, for all $j\in I$,
$s\in\Bbb Z$. Hence, we may assume that
$$\phi_{j,s}^{{}\pm{}}. v =\Phi_{j,s}^{{}\pm{}} v\ \ \ \ \ \ (j\in I,\ s\in\Bbb
Z)$$
for some $\Phi_{j,s}^{{}\pm{}}\in\Bbb C$. In particular, $v$ is a common
eigenvector of $k_1$ and $k_2$, and since $v\in\hat{U}_i.v_{\bp}$, its weight
is clearly of the form $\lambda-t\alpha_i$ for some $t\in\Bbb Z$. Then,
$x_{j,r}^+.v=0$ for $j\ne i$ as well. This shows that $v$ is a $\uqgh$--highest
weight vector, which contradicts the irreducibility of $V(\bp)$.\qed\enddemo

\demo{Proof of 5.3} (a) By 5.2, $\hat{U}_i.v_{\bp}$ is irreducible as a
representation of $U_{q_i}(\hat{sl}_2)$. By 4.3, there exists $0\ne
v\in\hat{U}_i.v_{\bp}\cap V(\bp)_{\lambda-\alpha_i}$ such that $x_{i,0}^+.v =
0$. Clearly then $x_{j,0}^+.v =0$ for $j\ne i$   so
$m_{\lambda-\alpha_i}\!(V(\bp)) >0$.

(b) If $P_i=1$, the statement is clear,  since then $\lambda-r\alpha_i\notin
P^+$. Let $P_i\ne 1$ be such that the roots of $P_i$ form a $q_i$--segment.  If
$m_{\lambda-r\alpha_i}(V(\bp))\ne 0$, then, by 1.4, we see that there exists
$0\ne v\in V(\bp)_{\lambda-r\alpha_i}\cap \hat{U}_i.v_{\bp}$ such that
$x_{i,0}^+.v =0$.   But, by 5.2 and 4.2(a), $\hat{U_i}.v_{\bp}$ is irreducible
as a representation of $U_{q_i}({sl}_2)$, so this is impossible.\qed\enddemo

Before showing that 5.1(b) holds,  we show that if one of $P_1$, $P_2$ is equal
to 1, say $P_1$ without loss, and the roots of $P_2$ form a $q_2$-segment, then
$[V(\bp)]$ is minimal. For, suppose $\bq\in\Cal{P}^{\lambda}$ is  such that
$[V(\bq)]$ is minimal and $[V(\bq)]\preceq[V(\bp)]$. Then $Q_1 =1$, and since
5.1(a) is neccessary, the roots of $Q_2$ form a $q_2$--segment. But then, for
some $t\in\Bbb C^{\times}$, $\tau_t^*(V(\bq))\cong V(\bp)$ by 3.4, so
$[V(\bq)]=[V(\bp)]$ and $[V(\bp)]$ is minimal. This establishes 5.1 when $r_i
=0$ for some $i=1,2$.

{}From now on, we assume that $P_i\ne 1$, $i=1,2$, and that the roots of $P_i$
form a $q_i$--segment with centre $a_i\in\Bbb C^{\times}$, $i=1$, $2$. To
complete the proof of 5.1, we need the following two results.
Status: RO

\proclaim{Proposition 5.4} Let $\ung$ be of type $C_2$ or $G_2$, and let $\mu
=r_1\lambda_1+r_2\lambda_2\in P^+$. Assume that, if $Q_i\ne 1$, the roots of
$Q_i$ form a $q_i$--segment of length $r_i$ and centre $b_i$.

\noindent (a) If $Q_i =1$ for some $i=1,2$, then, for all $s_i>0$,
$$m_{\mu- s_i\alpha_i}\!(V(\bq)) = 0, \;\; m_{\mu-\alpha_1-\alpha_2}\!(V(\bq))
=0.$$

\noindent (b) Assume that $Q_i\ne 1$, $i=1,2$. Let $M$ be a highest weight
representation of $\uqgh$ with highest weight $\bq\in\calp^{\mu}$ such that
$$m_{\mu-\alpha_1-\alpha_2}\!(M) =0,\ \ \  m_{\mu-\alpha_i}\!(M) =  0,\tag8$$
for $i=1,2$. Then,
$$\frac{b_1}{b_2} =q^{d_1r_1+d_2r_2+2d_2-1} \;{\text{or}}\;\;
q^{-(d_1r_1+d_2r_2+2d_1 -1)}.\tag9$$

\noindent (c) Assume that  $Q_i\ne 1$ and define
$\bq^{(i)}\in\calp^{r_i\lambda_i}$ as follows:
$$ Q_j^{(i)} =\cases Q_i &\text{if $i=j$,}\\
1 &\text{if $i\ne j$.}\endcases$$
 Let $v_i$ be a $\uqgh$--highest weight vector in $V(\bq^{(i)})$, and let $M
=\uqgh.(v_i\ot v_j)\subset V(\bq^{(i)})\ot V(\bq^{(j)})$, $ i\ne j$.  Then
 $$m_{\lambda-\alpha_1-\alpha_2}\!(M) =0\ \ \text{iff $\frac{b_i}{b_j}
=q^{-(d_1r_1+d_2r_2+2d_i-1)}$}.$$
 \endproclaim

If $\rho:\uqgh\to End(V)$ is a representation, denote the representation
$\rho\circ\hat{\omega}$ by $\hat{\omega}^*(V)$.
\proclaim{Proposition 5.5} Let $\ung$ be of type $C_2$ or $G_2$. Let
$\mu=r_1\lambda_1+r_2\lambda_2\in P^+$, let $\bq=(Q_i)_{i\in
I}\in\Cal{P}^{\mu}$, and let
$$Q_i(u) =\prod_{r=1}^{r_i}(1-a_{i,r}^{-1}u)\ \ \ \ \  (i=1,\,2).$$
Define polynomials $\overline{Q}_i(u)$ by
$$\overline{Q}_i(u) =\prod_{r=1}^{r_i}(1-q_i^2a_{i,r}u).$$
Then, $V(\overline{\bq})\cong \tau_t^*\hat{\omega}^*(V (\bq))$ for some
$t\in\Bbb C^{\times}$.
\endproclaim

Assuming these propositions, we complete the proof of 5.1 as follows. Suppose
that $[V(\bp)]$ is minimal but that $a_1/a_2$ has neither of the values stated
in 5.1(b). By 5.3(b), $m_{\lambda-\alpha_i}\!(V(\bp)) =0$, so by 5.4(b),
$$m_{\lambda-\alpha_1-\alpha_2}\!(V(\bp)) > 0.$$
Choose $\bq=(Q_i)_{i\in I}\in\Cal{P}^{\lambda}$ such that the roots of $Q_i$
form a $q_i$-segment with centre $b_i$, where $b_1/b_2$ has one of the values
in  5.1(b). By 5.4(c),
$$m_{\lambda-\alpha_1-\alpha_2}(V(\bq)) =0.$$
Hence, $[V(\bq)]\ne [V(\bp)]$. If $m_{\mu}(V(\bq)) > 0$, $\mu\ne\lambda$,
then by 5.4(c), $\mu =\lambda-s_1\alpha_1 -s_2\alpha_2$ where $s_1,s_2 >0$, and
since
$\mu\ne\lambda-\alpha_1-\alpha_2$, we have $\mu <\lambda-\alpha_1-\alpha_2$.
Hence $\mu$ satisfies 3.8(ii), and $[V(\bq)]\prec [V(\bp)]$, contradicting
minimality of $[V(\bp)]$.

Conversely, suppose $\bp$ is such that conditions 5.1(a) and 5.1(b) are both
satisfied. Choose $\bq=(Q_i)_{i\in I}\in\Cal{P}^{\lambda}$ such that $[V(\bq)]$
is minimal and $[V(\bq)]\preceq [V(\bp)]$. Since conditions 5.1(a) and (b) are
necessary, the roots of $Q_i$ must form a $q_i$-segment with centre $b_i$, say,
where $b_1/b_2$ also has one of the values in 5.1(b). If $a_1/a_2 =b_1/b_2$,
then, by 3.4, $V (\bq)\cong \tau_t^*(V(\bp))$ for some $t\in\Bbb C^{\times}$,
and then $[V(\bq)] =[V(\bp)]$. On the other hand, if $a_1/a_2 =
q^{d_1r_1+d_2r_2+2d_2-1}$ (resp. $q^{-(d_1r_1+d_2r_2+2d_1 -1)}$) and $b_1/b_2
=q^{-(d_1r_1+d_2r_2+2d_1 -1)}$ (resp. $q^{d_1r_1+d_2r_2+2d_2 -1}$), then by
5.5, $V(\bq)\cong \hat{\omega}^*(V(\bp))$, and again $[V(\bq)] =[V(\bp)]$. In
both cases, $[V(\bp)]$ is minimal.

We continue to assume 5.4 and prove 5.5.

\demo{Proof of  5.5}
We first reduce to the case when $V(\bq)$ is fundamental. By 2.6, $V(\bq)$ is
isomorphic to the unique irreducible subquotient of
$$\bigotimes_{i\in I}\bigotimes_{r=1}^{r_i}V(\lambda_i,a_{i,r})$$
which contains a $\uqg$--subrepresentation isomorphic to $V(\mu)$, and hence
$\hat{\omega}^*(V(\bq))$ is isomorphic to the unique irreducible subquotient of
$$\hat{\omega}^*\left(\bigotimes_{i\in
I}\bigotimes_{r=1}^{r_i}V(\lambda_i,a_{i,r})\right)\cong \bigotimes_{i\in
I}\bigotimes_{r=1}^{r_i}\hat{\omega}^*(V(\lambda_i,a_{i,r}))\tag10 $$
which contains a $\uqg$-subrepresentation isomorphic to $V$ (the order of the
factors in the tensor product on the right--hand side of (10) is the reverse of
that on the left--hand side). It is clear that the unique such quotient of
$$\bigotimes_{i\in I}\bigotimes_{r=1}^{r_i}V(\lambda_i,ta_{i,r}^{-1})$$
is isomorphic to $(\tau_{t}^*)^{-1}(V(\overline{\bq}))$.

To prove the result in the fundamental case, note that, by 3.3,
$$\hat{\omega}^*(V(\lambda_i,a_i))\cong V(\lambda_i,\overline{a}_i),\ \ \ \ \ \
 (i=1,2),$$
for some $\overline{a}_i\in\Bbb C^{\times}$ (not necessarily the complex
conjugate of $a_i$). Assume that $a_1/a_2 =q^{-(3d_1+d_2-1)}$. By 5.9, if $v_i$
is a $\uqgh$-highest weight vector in $V(\lambda_i,a_i)$, and $M =\uqgh.(v_1\ot
v_2)$, then $m_{\lambda_1+\lambda_2-\alpha_1-\alpha_2}(M) =0$.
Clearly, $M' =\uqgh.(v_2\ot v_1)\subseteq \hat{\omega}^*(M)$, hence
$m_{\lambda_1+\lambda_2-\alpha_1-\alpha_2}(M') =0$. By 5.9 again,
$$\frac{\overline{a}_2}{\overline{a}_1} =q^{-(d_1+3d_2-1)}.$$
Hence,
$${q_1^2a_1}{\overline{a}_1} = {q_2^2a_2}{\overline{a}_2},$$
so the result follows from 3.4.\qed
\enddemo

\demo{Proof of 5.4(a)} We assume that $d_2 =1$.

The fact that $m_{\mu-s_i\alpha_i}\!(V(\bq)) =0$ follows from 5.3(b).
If $Q_1=1$ it is enough to notice that $\mu-\alpha_1-\alpha_2\notin P^+$.

If $Q_2=1$, we must consider separately the cases when $\ung$ is of type $C_2$
or $G_2$. Let $v_\bq$ be a $\uqgh$-highest weight vector of $V(\bq)$.

If $\ung$ is of type $C_2$, then $x_0^+.v_\bq$ has weight
$r_1\lambda_1-\alpha_1 -2\alpha_2$, which is Weyl group conjugate to
$r_1\lambda_1-\alpha_1\in P^+$. Hence, if $m_\nu(V(\bq))>0$ and $x_0^+.v_\bq$
has a non-zero component in a $\uqg$-subrepresentation of $V(\bq)$ of highest
weight $\nu$, then $\nu=r_1\lambda_1$ or $r_1\lambda_1-\alpha_1$. But,
$m_{r_1\lambda_1-\alpha_1}\!(V(\bq))=0$ by 5.3(b), so
$x_0^+.v_\bq\in\uqg.v_\bq\cong V(r_1\lambda_1)$. Similarly, if $v_\bq^*$ is a
$\uqgh$-lowest weight vector of $V(\bq)$, then
$x_0^-.v_\bq^*\in\uqg.v_\bq^*=\uqg.v_\bq$. It follows that $x_0^{{}\pm{}}$
preserve $\uqg.v_\bq$, and hence that $\uqg.v_\bq$ is a
$\uqgh$-subrepresentation of $V(\bq)$. This not only proves 5.4(a), but the
following stronger result:
\proclaim{Proposition 5.6} If $\ung$ is of type $C_2$, where $d_2=1$, and if
$r\in\Bbb N$, then $r\lambda_1$ has an affinization which is irreducible as a
representation of $\uqg$ (this necessarily represents the unique minimal
element of $\calq^{r\lambda_1}$).\qed\endproclaim
If $\ung$ is of type $G_2$, $x_0^+.v_\bq$  is obviously killed by $x_2^+$,
since $[x_2^+,x_0^+] =0$. Hence, if $x_0^+.v_{\bq}$ has a non--zero component
$w$ in a $\uqg$--subrepresentation of $V(\bq)$ isomorphic to
$V(r_1\lambda_1-\alpha_1-\alpha_2)$, then $x_2^+.w =0$. This implies that
$x_1^+. w\ne 0$ since  the weight of $w$ (which is the same as the weight of
$x_0^+.v_{\bq}$) is $(r_1-1)\lambda_1$ which is less than
$r_1\lambda_1-\alpha_1-\alpha_2$. But then $(r_1-1)\lambda_1+\alpha_1$, and
hence also its Weyl group conjugate $(r_1-1)\lambda_1 +\alpha_1+3\alpha_2$, is
a weight of $V(r_1\lambda_1-\alpha_1-\alpha_2)$. This is impossible because
$(r_1-1)\lambda_1+\alpha_1+3\alpha_2 > r_1\lambda_1 -\alpha_1-\alpha_2$.

The proof of 5.4(a) is now complete.
\enddemo

 We assume from now on that $Q_i\ne 1$, $i=1,2$. To prove (b) and (c) we shall
need  the following result.

\proclaim{Lemma 5.7} Let $M$ be a highest weight $\uqgh$--module with highest
weight $\bq\in\calp^{\mu}$, $\mu =r_1\lambda_1+r_2\lambda_2$, and highest
weight vector $m$. Assume that $m_{\mu-\alpha_i}\!(M) =0$. Then,

(a) for $i=1,2$, $r\in\Bbb Z$,

$$h_{i,r}.m = H_{i,r}m,\;\; x_{i,r}^-.m = X_{i,r}x_{i,0}^- .m,$$
for some $H_{i,r},\ X_{i,r}\in\Bbb C$. If, in addition, the roots of $Q_i$ form
a $q_i$--segment with centre $b_i$, then
$$H_{i,1} = q_i^{-1}b_i^{-1}[r_i]_{q_i},\ \ \ X_{i,1} =b_i^{-1}q_i^{r_i-1}.$$

(b)  Let $W\subseteq M$ be the linear span of $\{ x_1^-x_2^-.m,
x_2^-x_1^-.m\}$. Then, the following are equivalent:

(i) $ x_{1,1}^-x_{2,0}^-.m\in W$;

(ii) $x_{i,r}^-x_{j,s}^-.m\in W$ for all $i\ne j\in\{1,2\}$, $r,s\in\Bbb Z$;

(iii) $m_{\mu-\alpha_1-\alpha_2}(M) =0$.
\endproclaim

We assume 5.7 and complete the proof of 5.4.

\demo{Proof of 5.4(b)} Suppose that (8) is satisfied. By 5.7(b), we can write
$$x_{1,1}^-x_{2,0}^-.m=Cx_{1,0}^-x_{2,0}^-.m+Dx_{2,0}^-x_{1,0}^-.m,$$
where $C$, $D\in\Bbb C$. Applying $x_{1,0}^+$, $x_{2,0}^+$ and $x_{2,1}^+$,
respectively, to both sides of this equation, and using 5.7(a) and the
relations in 1.3, we find the following system of equations for $C$, $D$:
$$\align
q_1^{-a_{12}}(b_1^{-1}q_1^{r_1-1}[r_1]_{q_1}-[a_{12}]_{q_1}q_1^{r_1}b_2^{-1}
q_2^{r_2-1})&=C[r_1-a_{12}]_{q_1}+D[r_1]_{q_1},\\
[r_2]_{q_2}b_1^{-1}q_1^{r_1-1}&=C[r_2]_{q_2} +D[r_2-a_{21}]_{q_2},\\
b_1^{-1}b_2^{-1}q_2^{r_2-1}[r_2]_{q_2}q_1^{r_1-1}
=Cb_2^{-1}q_2^{r_2-1}[r_2]_{q_2} & + \\
Dq_2^{-a_{21}} &
(b_2^{-1}q_2^{r_2-1}[r_2]_{q_2}-[a_{21}]_{q_2}q_2^{r_2}b_1^{-1}q_1^{r_1-1}).
\endalign$$
A straightforward calculation shows that these equations are consistent only if
(9) holds.
\enddemo

\demo{Proof of  5.4(c)}  We prove this when $\ung$ is of type $C_2$, the $G_2$
case is similar. By 5.3(b), we  know that, for all $i,j$,
$m_{\mu-\alpha_j}\!(V(\bq^{(i)}) =0$. Hence, $m_{\mu-\alpha_i}(M) =0$ for
$i=1,2$ and, by 5.7,
proving that  $m_{\mu-\alpha_1-\alpha_2}\!(M) =0$ is equivalent to proving that
$[x_{j,0}^-, x_{i,1}^-]_q.(v_i\ot v_j) = (qx_{j,0}^-x_{i,1}^- -
q^{-1}x_{i,1}^-x_{j,0}^-).(v_i\ot v_j)$ is a linear combination of the
following two elements:
$$\align x_{i,0}^-x_{j,0}^-(v_i\ot v_j) &=  x_{i,0}^-.v_i\ot x_{j,0}^-.v_j +
q_i^{-r_i}v_i\ot x_{i,0}^-x_{j,0}^-.v_j,\tag11\\
 x_{j,0}^-x_{i,0}^-(v_i\ot v_j) &= x_{j,0}^- x_{i,0}^-.v_i\ot v_j +
q_i^{-1}x_{i,0}^-.v_i\ot x_{j,0}^-.v_j.\tag12\endalign$$

Note that, by the isomorphism $f$ in 1.3,
we have $x_0^+ = [x_{2,0}^-,[x_{2,0}^-, x_{1,1}^-]_q](k_1k_2^2)^{-1},$
from which one deduces that
$$[x_2^+,x_0^+] =
(-1)^{\delta_{i,2}}(q^2-q^{-2})[x_{j,0}^-,x_{i,1}^-]_q(k_1k_2)^{-1} .$$
Using 1.2 and 5.7(a), one finds that
$$\align [x_2^+,x_0^+](v_i\ot v_j)&= [x_2^+,x_0^+].v_i\ot k_2k_0.v_j+ v_i\ot
[x_2^+,x_0^+].v_j\\
=(-1)^{\delta_{i,2}}(q^2-q^{-2}) & ( b_i^{-1}q^{-d_i-d_jr_j}
x_{j,0}^-x_{i,0}^-.v_i\ot v_j - b_j^{-1}q^{-1}v_i\ot
x_{i,0}^-x_{j,0}^-.v_j).\endalign$$
It is easy to see that this element is a linear combination of the elements in
(11) and (12) if and  only if $$\frac{b_i}{b_j} =q^{-(d_1r_1+d_2r_2+2d_i
-1)}.\qed$$
\enddemo

Finally, we give the

\demo{Proof of 5.7}
(a) That $m$ is a common eigenvector of the $h_{i,r}$ is  a consequence of the
fact that $m$ is the highest weight vector of $M$ and of  the relation between
the $\phi_{i,r}^{{}\pm{}}$ and the $h_{i,r}$ given in 1.3. If the roots of
$Q_i$ form a $q_i$--segment, then, by using (5), it follows immediately that
the eigenvalue of $h_{i,1}$ is as given. To see that $x_{i,r}^-.m$ is a
multiple of $x_{i,0}^-.m$,
it suffices to note that, in view of the relations in 1.3,
$x_{i,0}^+$ kills a suitable linear combination of $x_{i,r}^-.m$ and
$x_{i,0}^-.m$.

(b) By 5.7(a), to prove that (i) implies (ii), it suffices to prove that
$x_{i,r}^-x_{j,0}^-.m\in W$ for all $r\in\Bbb Z$, $i\ne j$. By the following
relation in $\uqgh$, and an obvious induction on $r$, we may assume that $i=1$,
$j=2$:
$$x_{2,r}^-x_{1,s}^- -q^{-a_{21}}x_{1,s}^-x_{2,r}^- =
 q^{-a_{21}}x_{2,r-1}^-x_{1,s+1}^- -x_{1,s+1}^-x_{2,r-1}^-.$$
Since we are given that $x_{1,1}^-x_{2,0}^-.m\in W$, we are reduced to proving
the following statement:
$$\text{if $x_{1,r}^-x_{2,0}^-.m\in W$ for some $r\in\Bbb Z$, then
$x_{1,r\pm1}^-x_{2,0}^-.m\in W$.}\tag13$$

To prove (13),  note first that  the relation
$$[h_{i,r},x_{j,s}^-] = -\frac{1}{r}[ra_{ij}]_{q^i}c^{|r|/2}x_{j,r+s}^-,$$
in $\uqgh$, and the fact that $c^{1/2}$ acts as the identity on $M$, imply that
there exist elements $H_r\in\uqgh$, $r\in\Bbb Z$,  which are linear
combinations of $h_{i,\pm r}, i=1,2$, such that, for all $m'\in M$, $r,s\in\Bbb
Z$,
$$ [H_s,x_{i,r}^-].m' =\delta_{i1}x_{i,r+s}^-.m'.$$
Now, (13) will follow if we prove that $H_{{}\pm 1}.W\subseteq W$. For $H_1$,
this follows from 5.7(a) and the assumption  that $x_{1,1}^-x_{2,0}^-.m\in W$.
For $H_{-1}$, it suffices similarly  to prove that $x_{1,-1}^-x_{2,0}^-.m\in
W$. By assumption,  we can write
$$x_{1,1}^-x_{2,0}^-.m = Ax_{1,0}^-x_{2,0}^-.m +Bx_{2,0}^-x_{1,0}^-.m, \tag14$$
for some $A,B\in\Bbb C$. If $A\ne 0$, applying $H_{-1}$\/ to both sides of (14)
gives the desired conclusion. If $A=0$, we use the same argument with $H_{-2}$
to get $x_{1,-1}^-x_{2,0}^-.m\in W$.

To prove that (ii) implies (iii), note that, by 1.4,
$M_{\mu-\alpha_1-\alpha_2}$ is spanned by
$\{x_{i,r}^-x_{j,s}^-.m\}_{i,j=1,2,r,s\in\Bbb Z}$. Hence,
$M_{\mu-\alpha_1-\alpha_2}=W$. Since $W\subseteq U_q(\ung).m\cong V(\mu)$,
$m_{\mu-\alpha_1-\alpha_2}(M)=0$.

For (iii) implies (i), suppose for a contradiction that
$x_{1,1}^-x_{2,0}^-.m\notin W$. Then $dim(M_{\mu-\alpha_1-\alpha_2})\ge 3$.
But, $dim(V(\mu)_{\mu-\alpha_1-\alpha_2})=2$. Since $m_{\mu-\alpha_i}(M)=0$ for
$i=1,2$, we must have $m_{\mu-\alpha_1-\alpha_2}(M)>0$.
\enddemo

\vskip36pt\noindent{\bf References}
\vskip12pt\noindent
1. Beck, J., Braid group action and quantum affine algebras, preprint, MIT,
1993.

\noindent 2. Chari, V. and Pressley, A. N., New unitary representations of loop
groups, Math. Ann. {\bf 275} (1986), 87-104.

\noindent 3.  Chari, V. and Pressley, A. N., Quantum affine algebras, Commun.
Math. Phys. {\bf 142} (1991), 261-83.

\noindent 4.  Chari, V. and Pressley, A. N., Small representations of quantum
affine algebras, Lett. Math. Phys. {\bf 30} (1994), 131-45.

\noindent 5.  Chari, V. and Pressley, A. N., {\it A Guide to Quantum Groups},
Cambridge University Press, Cambridge, 1994.

\noindent 6.  Chari, V. and Pressley, A. N., Quantum affine algebras and their
representations, preprint, 1994.

\noindent 7. Drinfel'd, V. G., A new realization of Yangians and quantized
affine algebras, Soviet Math. Dokl. {\bf 36} (1988), 212-6.

\noindent 8. Frenkel, I.B. and Reshetikhin, N. Yu., Quantum affine algebras and
holonomic difference equations, Commun. Math. Phys. {\bf 146} (1992), 1-60.

\noindent 9. Lusztig, G., {\it Introduction to Quantum Groups}, Progress in
Mathematics 110, Birkh\"auser, Boston, 1993.

\enddocument